\documentclass[grl]{agu2001} 
\usepackage{graphicx}

\authorrunninghead{KONTAR and P{\'E}CSELI}
\titlerunninghead{Enhanced ion acoustic lines}

\received{}
\revised{}
\accepted{}

\paperid{}

\authoraddr{{\it Permanent addresses:} E.~P. Kontar, Kelvin Building,
  University of Glasgow, Glasgow G12
  8QQ,  UK (eduard@astro.gla.ac.uk); H.~L. P{\'e}cseli,
University of Oslo, Physics Department, Blindern, N-0316 Oslo,
Norway (hans.pecseli@fys.uio.no)}

\begin{document}

\setkeys{Gin}{draft=false}

\title{Nonlinear wave interactions as a model for naturally enhanced ion
  acoustic lines in the ionosphere}

\author{E.~P. Kontar \& H.~L. P{\'e}cseli}

\affil{Centre for Advanced Study at the Academy of Sciences \&
Letters, N-0271 Oslo, Norway}

\begin{abstract}
Incoherent radar scatter from the ionosphere will, for equilibrium
conditions, show two symmetric ion-acoustic lines, one for each
direction of wave propagation. Many observation, from the EISCAT
Svalbard Radar (ESR) for instance, demonstrate that the symmetry
of this ion line can be broken, accompanied by an enhanced,
nonthermal, level of fluctuations, i.e.,~Naturally Enhanced
Ion-Acoustic Lines (NEIALs). Several models have been proposed
for explaining these naturally enhanced lines.
Here, we consider one of these, suggesting that decay of
electron beam excited Langmuir waves gives rise to enhanced
asymmetric ion lines in the ionosphere. We use a
weak-turbulence approximation, and identify crucial parameters for
Langmuir decay processes to be effective in generating the
observed signals.
\end{abstract}

\begin{article}

\section{Introduction}

Incoherent scatter radars are a standard tool for investigating
the properties of the Earth's ionosphere, and in the case of  thermal
equilibrium, the backscatter can be analyzed in terms of the
fluctuations-dissipation theorem from basic thermodynamics and
statistical mechanics
\citep{sedgemore-schulthess_stmaurice_2001}. Very
often it is found, however, that the ionospheric plasma is out of
equilibrium, and that the ion-line signal is distorted
\citep{buchert_et_al_1999,grydeland_et_el_2003}. Several models
can account for this feature: the symmetry of the natural ion-line
is broken, for instance, if a current is flowing through the
plasma \citep{sedgemore-schulthess_stmaurice_2001}, or if an
electron beam enhances electron plasma waves (Langmuir waves)
significantly above the thermal level, so that ion acoustic waves
are excited by parametric decay
\citep{tsytovich_1995}. Such models were invoked by
\cite{forme_1999} and \cite{forme_ogawa_buchert_2001}.

Although the observations as such are unambiguous
\citep{buchert_et_al_1999}, the interpretation is made difficult
by several problems: the observed features are often sporadic in
nature and can vary with time as well as altitude, on second and
km scales, respectively
\citep{sedgemore-schulthess_stmaurice_2001}. Also, we bear in mind
that the radar is usually obtaining backscatter at one selected
wave-vector ${\bf k}_R$, which is related to the scattering
wave-vector ${\bf k}_B$ by selection rules, which for a
mono-static radar gives ${\bf k}_B=2{\bf k}_R$. For the ESR-radar
\citep{buchert_et_al_1999}, we have a transmitter frequency of 500
MHz, giving $\lambda_B\equiv 2\pi/k_B=0.299$ m. For an altitude
$\sim 400$ km, with electron temperatures of $T_e\approx 3000$ K,
and plasma densities of $n_e\approx 2\times 10^{11}$ m$^{-3}$, we
find Debye lengths $\lambda_{De}\approx 8.5\times 10^{-3}$ m,
i.e.~$\lambda_B/\lambda_{De}\sim 35$, or $k_B\lambda_{De}\approx
0.18$.  In order to observe the ion sound wave we should have the
sound wavelength $\lambda_s$ equal to $\lambda_B$, implying that
the primary Langmuir wavelength $\lambda_L$ is approximately
$\frac{1}{2}\lambda_B$. This means that in a decay process, we
cannot usually, at a given altitude, expect to observe the first
generation Langmuir waves simultaneously with the sound waves
forming the low frequency part of the decay products. In the case
where we have a ``cascade'' of decaying waves
\citep{tsytovich_1995}, we might observe one or the other of the
decay products, and it might very well be the second or third
generation that is observed, instead of the first one. For the
parameters mentioned before we have $k_L\lambda_{De}\approx 0.09$.
The plasma conditions are strongly variable \citep{fontaine_2002}
and it is not always obvious under which conditions the enhanced
ion-lines are observed.

Several models have been proposed
\citep{rietveld_collis_stmaurice_1991,wahlund_et_al_1992}, but the
decay model may be the most promising, at least for explaining
some observed features
\citep{sedgemore-schulthess_stmaurice_2001,grydeland_et_el_2003}.
One problem with the Langmuir decay model seems to be that very
short wavelength primary Langmuir waves are sometimes needed to
account for the observations, below a few tens of Debye lengths,
$\lambda_{De}$. It might be possible to find a low velocity
electron beam which generates unstable waves for $v < 4v_{Te}$,
but the decay Langmuir wave (``daughter wave'') obtained from
these will be strongly Landau damped, implying that the growth
rate of the decay instability becomes negligible. Also, the
effects connected with plasma inhomogeneity have been largely
ignored, although a consistent treatment of the ionospheric plasma
density gradient can bring new understanding of the observational
results.

We discuss here the decay model in detail, including the most
important additional effects, such as the nonlinear Landau damping
and the effect of a large scale density gradient, to demonstrate a
set of criteria for the decay mechanism to be a viable explanation
for the enhanced ion-acoustic waves. These processes are here
modeled by a self-consistent set of dynamic equations, which are
solved numerically. A self-consistent description of beam-plasma
interactions, Langmuir wave scattering off ions (nonlinear Landau
damping), and plasma inhomogeneity can bring qualitatively new
results.

\section{Beam-driven Langmuir turbulence}

The basic equations are here solved for an initial value problem, which is
standard for these types of problems. This
gives a simplified alternative to the full problem, while  at
the same time retaining the important physics. Qualitatively, we might
transform temporal evolutions to spatial evolutions, using the beam velocity
for the transformation, i.e.,~$x\sim t v_b$. The analysis is
restricted to one spatial dimension for practical computational
reasons. We do not expect this to pose any serious limitation, in
part also because the growth rate for the decay instability is
known to have a maximum for aligned wave-vectors
\citep{thornhill_haar_1978}. Also, for the observations relevant
to the present study \citep{grydeland_et_el_2003} the radar beam
is basically directed along the magnetic field lines. Using the
assumptions mentioned above, we can write the system of kinetic
equations of weak turbulence theory
\citep{vedenov_velikhov_sagdeev_1962,kontar_2001_solar,kontar_pecseli_2002}
\begin{equation}
\frac{\partial f}{\partial t}= \frac{4\pi^2 e^2
}{m^2}\frac{\partial}{\partial v} \frac{W_k}{v}\frac{\partial
f}{\partial v},\;\;\; \label{eqk1}
\end{equation}
where $f(v,t)$ is the spatially averaged electron distribution
function. $W(k,t)$ is the spectral energy density of Langmuir
waves and plays the same role for waves as the electron
distribution function for particles. This equation accounts for
the quasi-linear relaxation of the electron beam.

For an inhomogeneous plasma, with a density gradient along the
direction of beam propagation ${\cal L}=\omega _{pe}/(\partial
\omega _{pe}/\partial x)^{-1}$, there are two distinct processes
affecting Langmuir waves \citep{kontar_2001_pphcf}. First, the
local plasma frequency $\omega _{pe}$ is spatially dependent
leading the instability growth-rate to change with distance.
Second, and more important, a small spatial movement of Langmuir
waves with group velocity $\partial \omega _L/\partial k<<v_{Te}$
leads to a substantial change in the wavenumber of the plasma
waves, since $\omega _L (k,x)\approx \omega_{pe}(1+\Delta n/n
+\frac{3}{2}(k_L\lambda_{De})^2)$. By taking the wave frequency
constant along the ray path for a stationary density profile, we
find, as an estimate, the variation of the Langmuir wavenumber to
be $(k_{L1}\lambda_{D1})^2\approx
(k_{L2}\lambda_{D2})^2+\frac{2}{3} \Delta n/n$, where we follow
the wave from positions labeled 1 to 2, having a density
difference of $\Delta n$. This turns out to be an effective
mechanism for transferring wave energy from large to small
wavenumbers or vice-versa \citep{kontar_pecseli_2002}, depending
on the sign of $\Delta n$. This process has seemingly been ignored
in previous studies of the enhanced ion-lines.

For the small density variation $\Delta n/n <(k\lambda _{De})^2$,
we can ignore the frequency variations, while
retaining the wavenumber shift due to inhomogeneity in the equation
for the evolution of Langmuir waves in the WKB (geometrical optics)
approximation
\begin{eqnarray}
&&\frac{\partial W_k}{\partial t}-\frac{\omega_{pe}}{\cal L}
\frac{\partial W_k}{\partial k} =\frac{\pi
\omega_{pe}^3}{nk^2}W_k\frac{\partial f}{\partial
v}-2\nu_{ce} W_k \nonumber\\
&&\hspace{2.5cm}+St_{decay}(W_k,W^s_k)+St_{ion}(W_k)\, ,
\label{eqk2}
\end{eqnarray}
where $-({\omega_{pe}}/{\cal L}) \, {\partial W_k}/{\partial k}$
accounts for the effective acceleration
$-\partial\omega_{pe}/\partial x$ of a Langmuir wave packet by
a density gradient. This term is well known, for instance, from
geometrical optics, and is caused by the spatial variation of the
index of refraction. This acceleration can imply that
waves excited by the beam can reach phase
velocities above the beam velocity, after which they
will be re-absorbed by Landau damping.

The system (\ref{eqk1}) and (\ref{eqk2}) self-consistently
describes the resonant interaction $\omega_{pe}=kv$ of electrons
and Langmuir waves. The first term on the right hand side of
(\ref{eqk2}) accounts for the waves generated by the electron
beam. A simple damping term with collision frequency $\nu_{ce}$
accounts for the collisional damping of the waves.  We included
the collisional damping of Langmuir waves with $\nu _{ce}=\nu
_{ei}+\nu _{en}$, where $\nu _{ei}=(34+4.18\ln
(T_e^3/n_e))n_eT_e^{-3/2}$ is for electron-ion collisions and $\nu
_{en}=5.4\times 10^{-10}n_{gas}T_e^{1/2}$ accounts for electron
neutral collisions  \citep{itikawa_1973}. The former dominates the
latter at higher altitudes while electron-neutral collisions are a
faster process below the F region maximum.

The third term $St_{decay}$ accounts for the change of wave energy
due to decay processes $L\rightarrow L'+S$, with $L$ and $S$
  denoting Langmuir and ion sound waves, respectively. We introduced also
$St_{ion}$ to account for the nonlinear Landau damping, which is
also known as scattering off ions, $L+i\rightarrow L'+i'$
\citep{tsytovich_1995}. These processes are efficient in
redistributing energy within the spectra. The nonlinear processes
accounted for by $St_{decay}$ and $St_{ion}$ are effective in the
scattering of primary, beam generated waves with wavenumber ${\bf
k}$, into secondary Langmuir waves with wavenumber ${\bf
k}^{\prime}\approx {\bf -k}$.  The decay of Langmuir waves
leads to generation of the ion-sound waves with ${\bf k}_s\approx
2{\bf k}$.

The time evolution of ion acoustic fluctuations,  with
a spectral energy $W^{s}_{k}$, is described by
\begin{eqnarray}
&&\frac{\partial W^{s}_{k}}{\partial t}=-2\gamma^s_k
W^{s}_{k}\nonumber\\
&&\hspace{.25cm}-\alpha {\omega^s_k}^2\int\left(
\frac{W_{k^{\prime}-k}}{\omega_{k^{\prime}-k}}\frac{W^s_{k}}{\omega^s_{k}}-
\frac{W_{k^{\prime}}}{\omega_{k^{\prime}}}\left(\frac{W_{k^{\prime}-k}}{\omega_{k^{\prime}-k}}+
\frac{W^s_{k}}{\omega^s_{k}}\right)\right)\nonumber\\
&&\hspace{.5cm}\times\delta
(\omega_{k^{\prime}}-\omega_{k^{\prime}-k}-\omega^s_k)dk^{\prime} \, ,
\label{eqk3}
\end{eqnarray}
with
\begin{eqnarray}
&&St_{decay}(W_k,W^s_k)=\nonumber\\
&&\alpha\omega_{k}
\int\omega^s_{k^{\prime}}\left[ \left(
\frac{W_{k-k^{\prime}}}{\omega_{k-k^{\prime}}}\frac{W^s_{k'}}{\omega^s_{k'}}-
\frac{W_k}{\omega_k}\left(\frac{W_{k-k^{\prime}}}{\omega_{k-k^{\prime}}}+
\frac{W^s_{k^{\prime}}}{\omega^s_{k^{\prime}}}\right)\right)\right.
\nonumber\\
&&\hspace{.5cm}\times \,\delta
(\omega_{k}-\omega_{k-k^{\prime}}-\omega^s_{k^{\prime}})-\nonumber\\
&&\hspace{1.0cm}\left.\left(
\frac{W_{k+k^{\prime}}}{\omega_{k+k^{\prime}}}\frac{W^s_{k'}}{\omega^s_{k'}}-
\frac{W_k}{\omega_k}\left(\frac{W_{k+k^{\prime}}}{\omega_{k+k^{\prime}}}-
\frac{W^s_{k^{\prime}}}{\omega^s_{k^{\prime}}}\right)\right)\right.
\nonumber\\
&&\hspace{2.5cm}\times \left.\delta
(\omega_{k}-\omega_{k+k^{\prime}}+\omega^s_{k^{\prime}})\right]
dk^{\prime}\, ,
\end{eqnarray}
%
\begin{eqnarray}
&&St_{ion}(W_k)=\nonumber\\
&&\hspace{-.25cm}\beta \omega_k\int
\frac{(\omega_{k^{\prime}}-\omega_k)}{v_{Ti}|k-k^{\prime}|}
\frac{W_{k^{\prime}}}{\omega_{k^{\prime}}} \frac{W_k} {\omega_k}
\exp\left[-\frac{(\omega_{k^{\prime}}
-\omega_k)^2}{2v_{Ti}^2|k-k^{\prime}|^2}\right]
dk^{\prime} \, ,
\end{eqnarray}
where
\begin{equation}
\alpha=\frac{\pi \omega^2_{pe}(1+3T_i/T_e)}{4n\kappa T_e},\;\;\;
\beta=\frac{\sqrt{2\pi}\omega^2_{pe}}{4n\kappa T_i(1+T_e/T_i)^2},
\end{equation}
\begin{equation}\label{gam_sk}
\hspace{-.5cm}\gamma^s_k=\sqrt{\frac{\pi}{8}}\omega^s_k\left[\frac{v_s}{v_{Te}}
+\left(\frac{\omega
^s_k}{kv_{Ti}}\right)^3\exp\left[- \left(\frac{ \omega ^s_k}{
kv_{Ti} }\right)^2 \right]\right] \, .
\end{equation}
Here $\gamma^s_k$ is the Landau damping rate of ion-sound waves,
containing both electron and ion components, $v_s=\sqrt{\kappa
T_e(1+3T_i/T_e)/m_i}$ is the sound speed, while $W(k,t)$ and
$W^s(k,t)$ are the spectral energy densities of Langmuir waves and
ion-sound waves, respectively. It turns out that for parameters
relevant here, Landau damping of ion-sound waves dominates
collisional damping. The relative importance of the various terms
in (\ref{eqk3}) can be studied by including them in the numerical
calculations one at a time. Since the equations are non-linear,
this procedure can give an estimate only.

\begin{table}\label{table1}
\caption{Varying parameters}
\begin{tabular}{cccccc}
Altitude, km &  $T_e$, K & $T_i$, K & $f_{pe}$, MHz  &
$n_{gas}$,cm$^{-3}$&
${\cal L}$, km \\
400  &  3100 &  1400 &  4.0   &    4$\times 10^{8}$  &  $\infty$\\
400  &  3100 &  1400 &  4.0   &    4$\times 10^{8}$   &  $+30$\\
250  &  2200 &  1100 &  5.0   &    4$\times 10^{9}$  &   $-30$\\
\end{tabular}
\end{table}

\begin{figure}
\noindent
\begin{center}
\includegraphics[width=73mm]{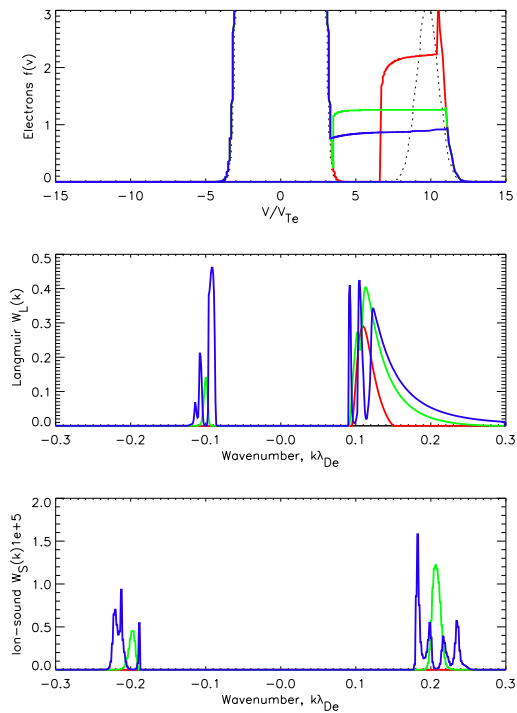}
\end{center}
\caption{Electron distribution function $f(v)v_b/n_b$,
Langmuir wave, $W_k\omega _{pe}/mn_bv_b^3$, and ion-sound wave,
$W^s_k \omega_{pe}/mn_bv_b^3$, spectra at four different times,
$t=0.0$ ms (black dashed-line), $t=0.2$ ms (red line), $t=1.5$ ms
(green line), $t=3.0$ ms (blue line). See also Figure
\ref{fig:FLS_hom}.} \label{fig:FLS_hom_sp}.
\end{figure}

\section{Numerical Results}

The set of equations (\ref{eqk1})--(\ref{gam_sk}) were solved
numerically using the method described by
\cite{kontar_pecseli_2002}, here for downward propagating
electron beams.

Common parameters for the numerical analysis are $n_{b}=20$
cm$^{-3}$, $v_{b} =3\times 10^8$ cm/s ($\sim 26$~eV), with the
subscript $b$ indicating ``beam''. As a reference parameter set we
take those listed in Table~\ref{table1}. Relevant electron-neutral
collision frequencies can be deduced and extrapolated from known
results, as summarized by, for instance, \citet{itikawa_1973}.
Since the dominant ion in the ionosphere at 250-400km is O$^+$,
the effective ion mass $m_i$ was taken to be 16 proton masses. We attempted
to use representative plasma parameters, so that the numerical results can
be used for qualitative as well as quantitative comparisons with
observations.

\begin{figure}
\noindent
\includegraphics[width=85mm]{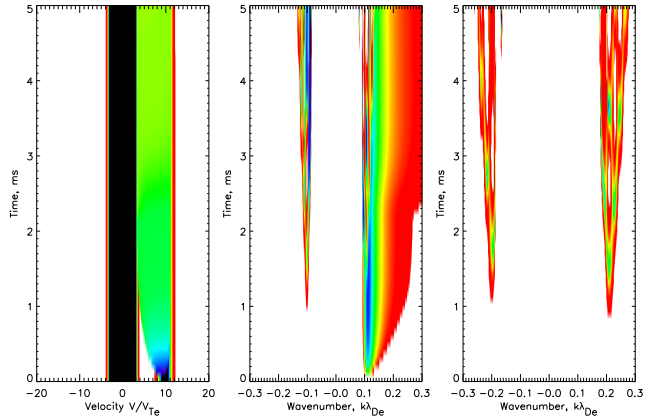}
\caption{Temporal and spectral evolution of electron beam (left
panel), Langmuir waves (middle panel), and ion-sound waves (right
panel) for homogeneous plasma and parameters at altitude 400km.
White-red-green-blue-black indicate growing intensity,
respectively. The background Maxwellian distribution appears as a black
band in the left panel.} \label{fig:FLS_hom}
\end{figure}

\begin{figure}
\noindent
\includegraphics[width=85mm]{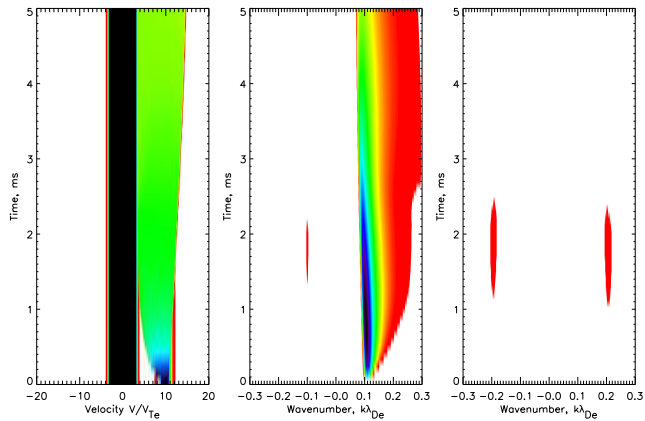}
\caption{The same as Figure \ref{fig:FLS_hom}, but for increasing
plasma density along the beam propagation with parameters for 400~km
altitude (above the F maximum).} \label{fig:FLS_inhom}
\end{figure}

\begin{figure}
\noindent
\includegraphics[width=85mm]{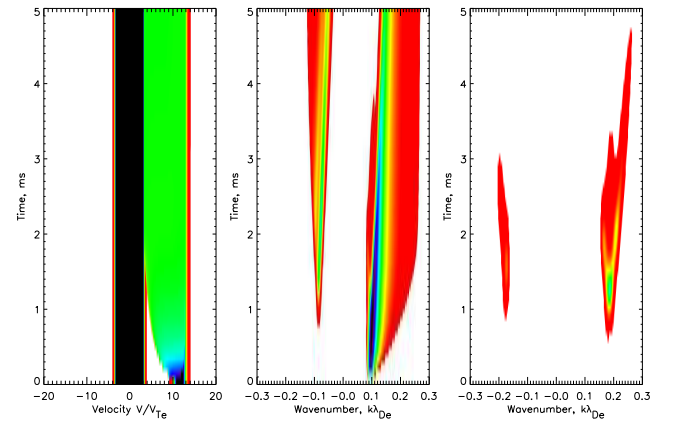}
\caption{The same as Figure \ref{fig:FLS_hom}, but for decreasing
plasma density with parameters for 250~km altitude (below the F maximum).}
\label{fig:FLS_250}
\end{figure}

Figures~\ref{fig:FLS_hom_sp} and~\ref{fig:FLS_hom} show the
evolution of beam-plasma system in a homogeneous plasma. The
distributions of electrons, Langmuir and ion-sound waves are shown
for various times. The most rapid process in the system is the
quasi-linear relaxation, followed by wave decay, while here the
scattering off ions plays only a secondary role. The turbulence
exhibits fine structure both in the Langmuir and ion-sound wave
spectra (Figure \ref{fig:FLS_hom_sp}). This structure appears due
to the decrease of the absolute value of $k$ at every decay by
$\Delta k_d\lambda_{De}=2\sqrt{m_e/m_i}\sqrt{1+3T_i/T_e}/3$ and
scattering off ions by $\Delta k_s\lambda
_{De}=2\sqrt{m_e/m_i}\sqrt{T_i/T_e}/3$.

The overall picture changes dramatically when plasma inhomogeneity
is accounted for (Figure \ref{fig:FLS_inhom}). For the parameters
in Figure \ref{fig:FLS_inhom} we have $\Delta k_d\lambda
_{De}\approx 0.005$ and $\Delta k_s\lambda _{De}\approx 0.002$,
and for the density gradient ${\cal L}=30$~km the wavenumber shift
rate is $dk/dt\sim \omega_{pe}/{\cal L}\sim
10$~cm$^{-1}$~s$^{-1}$. As an electron beam propagates downwards
above the F maximum, the generated Langmuir waves experience a
decrease in $k$, facilitating further decay of Langmuir waves with
the same $k$ (Figure \ref{fig:FLS_inhom}). {\em Below} the
F-region maximum, the wavenumber decrease due to decay and
scattering can be compensated by the density inhomogeneity for
approximately $0.5$~ms. Similarly, an increasing plasma density
can compensate an increase in $k$ along the beam direction in
Figure \ref{fig:FLS_250}.

\section{Discussions and Conclusions}

{\em Density gradient:} The steady state plasma distribution has a
``natural'' large scale density gradient, which changes sign when
going from below to above the F-region maximum. The scale length
of this gradient is some tens of kilometres. In addition, we might
expect that under disturbed conditions, we might have gradients with shorter
scale-lengths, where the direction of these can have both
signs. The shift of Langmuir waves in $k$-space is strong enough
to produce substantial changes in local Langmuir turbulence
spectra as illustrated in Figures \ref{fig:FLS_inhom} and
\ref{fig:FLS_250}. The results in these figures demonstrate that an
interpretation of NEIALs in terms of Langmuir decay
\citep{forme_1999,forme_ogawa_buchert_2001} is most convincing in regions
with vanishing or negative density gradients, corresponding to altitudes
around or below the F maximum.

{\em Scattering off ions:} The decay of Langmuir waves into
ion-acoustic and back-scattered Langmuir waves is normally most
efficient for plasmas with $T_e>>T_i$. The growth rate of the
decay or parametric instability should be higher than the
ion-acoustic damping rate, producing a threshold for the
instability. The damping rate is strongly dependent on $T_e/T_i$,
which can be small for the lower parts of the F-region, while scattering
off ions has no such threshold and dominates for $T_e/T_i \sim 1$.
The enhanced level of Langmuir waves with $k'_L\approx -k_L$ makes
the term $W_kW_{k-k'}$ in (\ref{eqk3}) efficient and thus
stimulates the decay. In  contrast to  a pure parametric
instability with a single shoulder, it gives a close to simultaneous
growth of two ion-acoustic shoulders, as in Figure~\ref{fig:FLS_inhom}. In
this case the radar backscatter has the best possibility of indicating an
enhancement of {\em both} ion lines.

We found that for altitudes above $\sim 150$ km, the quasi-linear
beam relaxation due to the collective effects is faster than that
caused by collisions. The decay instability, the scattering off
ions, and the shift due to a plasma inhomogeneity act on a longer
time-scale. It can be seen in Figures~\ref{fig:FLS_hom_sp} and
\ref{fig:FLS_hom} that the decay leads to antisymmetric structures
associated with multiple decay in the spectra of Langmuir
turbulence.

We have also considered the case where the background plasma is
nonthermal, here simulated by using a superposition of two
Maxwellians with different temperatures, consistent with
observations \citep{guio_lilensten_1999}. We have found the most
significant effect of a small density (1\% or less) warm electron
component (temperature $\sim 20\times$ the background temperature
$T_e$) to be a change in the Landau damping of the backscattered
Langmuir wave. When this wave is damped, the ion signal becomes
weaker. The present results were derived for {\em downwards}
propa\-gating electron beams, but qualitatively, the results
apply for upgoing beams as well, if the sign of ${\cal L}$ is changed.

\begin{acknowledgments}
The present
work was carried out as a  part of the project ``Turbulence in Fluids and Plasmas'',
conducted at  the Centre
for Advanced Study (CAS) in Oslo in 2004/05. We also thank Anja Str{\o}mme for valuable discussion in
the early stages of this work. Discussions with Jan Trulsen are
appreciated.
\end{acknowledgments}


\end{article}
\end{document}